\documentclass[%
 reprint,
 amsmath,amssymb,
 aps,
 prl
]{revtex4-2}

\usepackage{multirow}
\usepackage{graphicx}
\usepackage{dcolumn}
\usepackage{bm}
\usepackage{comment}

\usepackage{color}
\begin{document}

\title{Contribution  of collapsars, supernovae, and neutron star mergers to the evolution of r-process elements in the Galaxy}

\author{Yuta Yamazaki$^{1,2}$}\email{y.yamazaki@grad.nao.ac.jp}
\author{Toshitaka Kajino$^{1,2,3}$}
\author{Grant J. Mathews$^{2,4}$}
\author{Xiaodong Tang$^{5}$}
\author{Jianrong Shi$^{6}$} 
\author{Michael A. Famiano$^{2,7}$}

\affiliation{%
\\
$^1$Graduate School of Science, The University of Tokyo, Hongo, Bunkyo-ku, Tokyo 11-0033, Japan\\
$^2$National Astronomical Observatory of Japan, 2-21-1 Osawa, Mitaka, Tokyo 181-8588, Japan
}%

\affiliation{
$^3$School of Physics, and International Research Center for
    Big-Bang Cosmology and Element Genesis, Beihang University, Beijing 100083, P. R. China\\
$^4$ Department of Physics and Center for Astrophysics, University of Notre Dame, Notre Dame, IN 46556, USA
}%

\affiliation{
$^5$Institute of Modern Physics, Chinese Academy of Science, Lanzhou, Gansu 730000, P. R. China
}%

\affiliation{
$^6$Key Laboratory of Optical Astronomy, National Astronomical Observatories, Chinese Academy of Science, Beijing 100012, P. R. China
}%

\affiliation{
$^{7}$Physics Department, Western Michigan University, Kalamazoo, MI 49008-5252 USA
}%
\date{\today}

\begin{abstract}
We study  the evolution of rapid neutron-capture process (r-process) isotopes in the Galaxy.   We analyze relative contributions from core collapse supernovae (CCSNe), neutron star mergers (NSMs) and collapsars under a range of astrophysical conditions and nuclear input data. Although the r-process in each of these sites can lead to similar (or differing) isotopic abundances, our simulations reveal that the early contribution of r-process material to the Galaxy was dominated by CCSNe and collapsar r-process nucleosynthesis, while the NSM contribution is unavoidably  delayed even under the assumption of the shortest possible minimum merger time.  
\end{abstract}

\maketitle

The origin by rapid neutron capture (r-process) of nearly  half of the heavy atomic nuclides from iron to uranium remains an open question \cite{kajino2019}. 
The neutrino-driven wind (NDW) of core collapse supernovae (CCSNe) is now believed to produce only light r-process elements~\cite{wanajo2013}. The magneto-hydrodynamically driven jet (MHDJ) from rapidly rotating, strongly magnetized CCSNe is an alternative site  \cite{winteler2012}. Also, there is  a growing consensus \cite{hotokezaka2018,frebel18} that neutron star mergers could be the dominant contributor to r-process elements in the Galaxy.   This is due in part to the discovery of gravitational waves from the binary neutron star merger (NSM) GW170817 and its associated kilonova  and GRB170817A~\cite{gw170817,grb170817a}.   There is also  evidence \cite{Ji16,roederer2018} for massive ejection of r-process material without associated supernova enrichment  in the dwarf galaxy Reticulum II.


 Of interest to this letter is that in addition to the above sources, a single massive star collapsing to a black hole (collapsar) may also be a viable  site for the main r-process abundances~\cite{siegel2019}. It has been argued \cite{cote2018} that extra production site for the r-process  may be required that was active in the early Galaxy but fades away at higher metallicity.  Otherwise, it is not possible to account for the decrease  of r-process elements with iron after the onset of Type I supernovae (at [Fe/H] $\sim -1$).  In this paper we identify this source with collapsars and closely examine the  galactic chemical evolution (GCE)  of r-process abundances under a  range of astrophysical conditions for each possible  including collapsars.  We conclude that collapsars provide the desired source.  Moreover, we show that even in the most optimistic circumstances,  NSMs can only contribute in a minor way to the present solar-system r-process abundances.

There is a fundamental difference between NSMs and CCSNe.  There is an unavoidable  time delay from formation of the progenitors until the ejection of  r-process material.
We demonstrate  that even in the limit that the shortest timescale for mergers is only  the stellar evolution time scale ($\sim 10^6$ yr) it is inevitable that neutron star binaries  are formed with a distribution of separation distances and merger timescales.   This leads to a delay in the arrival of r-process material in the solar neighborhood, and constrains the NSM contribution  to the solar-system abundances.     

SNe and collapsars  result when a single massive star that completes its evolution within a few Myr. They can enrich r-process elements in the interstellar medium (ISM) of the Galactic disk from the earliest times. On the other hand, NSMs involve the remnants of previously exploded massive stars. Their observed occurrence rate is $\sim 0.1-1 \%$ of the observed galactic SN rate.  The observed orbital properties of binary pulsars imply a coalescence timescale ranging from a few hundred Myr to longer than the Hubble time~\cite{Swiggum2015}.

The universal elemental abundance pattern observed in metal-poor halo stars and the solar system (universality)~\cite{sneden2008} suggests that  only a single site contributed to the the r-process elements~\cite{mathews1992,argast2004,hotokezaka2018,ishimaru2005,cote2018}. However, the isotopic abundance patterns can be quite different even though the  elemental $Z$-distributions are universally similar \cite{shibagaki2016,suzuki2018,siegel2019}.

This motivates the study described herein of the relative contributions of  multiple r-process sites (CCSNe, NSMs and collapsars)  and their  cosmic  evolution. We have  utilized  a widely employed  GCE model \cite{timmes1995} adapted to calculate r-process contributions from various sources  in the solar neighborhood.   We have  explored a large parameter space of r-process models, nuclear input, and astrophysical parameters to examine the general features of the evolution of r-process abundances. 

We find that CCSNe and collapsars must dominate the r-process abundances  in the early Galaxy, while NSMs can only arrive later when the metallicity has  already been enriched to -1$<$[Fe/H].  This  is independent of the minimum timescale   for binary neutron-star coalescence and is a consequence of the unavoidable distribution of binary separation distances at formation.  


The observed cosmic star formation rate  (SFR) \cite{madau2014} as well as SPH galactic chemo-dynamical evolution (GCDE) models of spiral galaxy evolution \cite{kobayashi2004} and dwarf spheroidal galaxy evolution \cite{hirai2015}  all indicate  that the SFR at first rises and then diminishes with time.  
The GCE model \cite{timmes1995} adopted in this study produces a SFR  similar to that observed and deduced from GCDE simulations. 
This GCE model also reproduces well the chemical evolution of light elements from hydrogen to zinc~\cite{timmes1995}. In this study we have extended this   model to include the r-process contributions from NDW, MHDJ, collapsars and NSM.



Gas evolution involves a cycle of star formation,  stellar evolution and nucleosynthesis; ejection of material into the interstellar medium (ISM); mixing of ejecta with the ISM; and  formation of the next generations of stars. We adopt an exponentially declining  galactic inflow rate  with timescale of 4 billion years consistent with the hierarchical clustering paradigm.
Although, the merger of dwarf galaxies into the galactic halo can bring some r-process enriched stars, the bulk of the inflowing gas consists of r-process depleted  material from the circumgalactic medium \cite{celine20}.

The Surface density $\sigma_i$ of isotope $i$ in the ISM then obeys,

	\begin{equation}\label{eq:r_evo}
	\begin{split}
	    \dot{\sigma}_{i}(t) 
        = 
        &\ \sum_\mu \  \epsilon_{\mu}\ \int_{m_l}^{m_h} \ E_{i,\mu}\  B(t-\tau(m)) \  \phi(m) \ dm \\[5pt]
	    &+\epsilon_{\rm NSM}\ \iiint ^{M_{h}} _{M_{l}}
         dq\ da
          \  P_a(a)\  P_q(q)\\[5pt]
        &\times E_{i,{\rm NSM}}\ B(t - \tau(m_2)  -  \tau_{g}(a)) \  \phi(M) \ dM\ \\[5pt]
	       & \ - \ B(t) \  \frac{\sigma_{i}}{\ \sigma_{gas}\ }, \\[5pt]
	   \end{split}
	\end{equation} 
where the first and second terms on the  r.h.s. describe the enrichment of newly produced  nuclei by explosive nucleosyntheses in different astrophysical sites, i.e~$\mu$ = NDW, MHDJ, Collapsars, and NSMs, respectively. The third term accounts for the loss from the ISM due to  star formation, where $\sigma_{gas}$ is the total gas surface density.
$E_{i,\mu}$ is the yield of isotope $i$ from each  astrophysical site $\mu$. The quantity $B(t)$ is the star formation rate, and $\phi(m)$ is the initial mass function which we adopt from \cite{kroupa2001}. 

For the present illustration we adopt  the abundance distribution of r-process nuclei in the NDW model from the  1.8$M_\odot$ proto-neutron star yields of~\cite{wanajo2013}. We adopt yields from the NSM model of~\cite{suzuki2018}, and the MHDJ model yields are  taken from~\cite{nishimura2012}. For the collapsar model, we adopt the yields of~\cite{nakamura2015,famiano2020}. 

In Eq.~(\ref{eq:r_evo}), the enrichment of r-process nuclei from supernovae and collapsars in the ISM is delayed by   the period $\tau({\rm m})$ from star formation to the death of the progenitor star of mass $m$.
We adopt lifetimes from ~\cite{schaller1992} for massive stars and from \cite{woosley1995} for stars with $m\le10$ M$_\odot$. In \cite{woosley1995} it was estimated that the metallicity effect on the progenitor lifetime is only about $\sim$5\% and is ignored.

The quantities $\epsilon_{\mu}$ and $\epsilon_{\rm NSM}$ are the fractions of stars resulting in each event in a given mass range $m_l\sim m_h$ or $M_l\sim M_h$ for a single or binary system. In this study, we take these as efficiency parameters, which are adjusted to the solar  r-process abundances.  We here define the abundance of solar r-process nuclei as the sum of nuclides in mass range $90\leq A\leq 209$.

In the present study we do not consider  the slight metallicity dependence of various contributions.  
For the NSM contribution in Eq.~(\ref{eq:r_evo}) we include the long gravitational wave coalescence time $\tau_{\rm g}$ in addition to $\tau({\rm m})$. We adopt a paradigm  whereby main-sequence binaries of total mass $M =m_1+m_2$ are formed within a gas cloud.  The heavier star with mass $m_1$ explodes first to form a neutron star or black hole, followed by the second CCSN of the lighter progenitor.  

Binary population synthesis studies indicate that the minimum  coalescence time is $\sim$ 100 My.  This is in reasonable agreement with the lower limit estimated from observed binary pulsars~\cite{Swiggum2015}. 
In the quadrupole limit, the coalescence time $\tau_{\rm g}$ from an initial binary separation $a$  scales as $a^4$. We ignore the dependence on eccentricity as tidal interactions should circularize the orbits.  

The event rate of NSMs  in Eq.(\ref{eq:r_evo}), includes  the uniform probability $P_q(q)$ for a given mass ratio of the secondary to total mass, $q\equiv m_2/M$, and the probability $P_a(a)\propto1/a$, of an  initial separation $a$.  This is consistent with the  observationally inferred   coalescence time, $P_{\tau_g}\propto{\tau_g}^{-1}$\cite{beniamini2019, simonetti2019}. The lower limit of the initial separation $a$ constrains the minimum coalescence time. 

The observed lower limit of the coalescence timescale of binary neutron stars is $\sim$50 My~\cite{Swiggum2015}.
However, there are several effects that might shorten coalescence time. If the binary system is perturbed by the third stellar object or if there exits a common envelope, these would accelerate orbital energy loss rate. We therefore treat $\tau_{\rm g}$ as a parameter, and allow shorter minimum coalesce times, i.e $\tau_{\rm g}$ = 1, 10 and 100 My in our GCE calculations.

We consider two combinations of multiple astrophysical sites for r-process nucleosynthesis:  One set of models  includes only CCSNe and NSM contributions to the  r-process.  The other includes the  collapsar contribution. 

We also considered two sets of nuclear physics input. One adopts a symmetric fission fragment distribution (FFD) and the other  an asymmetric FFD. The FFD strongly affects abundances near r-process peaks as discussed below. 
The  four models are labeled as follows:  Model sym- and asym- ignore the collapsar contribution, adopting symmetric and asymmetric FFDs respectively.
Model sym+ and asym+ include collapsars with symmetric and asymmetric FFDs respectively.



We take efficiency parameters $\epsilon_\mu$ in Eq. (1) as free parameters  adjusted so as to best fit the solar-system r-process abundances at the time of solar-system formation. 
The reduced chi square for the  fitted abundance pattern for each of  the four models is shown in Table 1.

The collapsar contribution improves the goodness of fit even after accounting for the extra degree of freedom.
The collapsar abundances do not exhibit an underproduction near $A=140-145$ which appears in most previous r-process calculations \cite{kajino2019}.
The r-process in collapsars undergoes fission recycling so many times that the fission fragments enhance the abundances at $A\approx130-150$.
This potentially solves the long standing problem in r-process nucleosynthesis of underproduction of nuclides with $A=140-145$.

\begin{figure}
\includegraphics
[width=1.0\linewidth]
{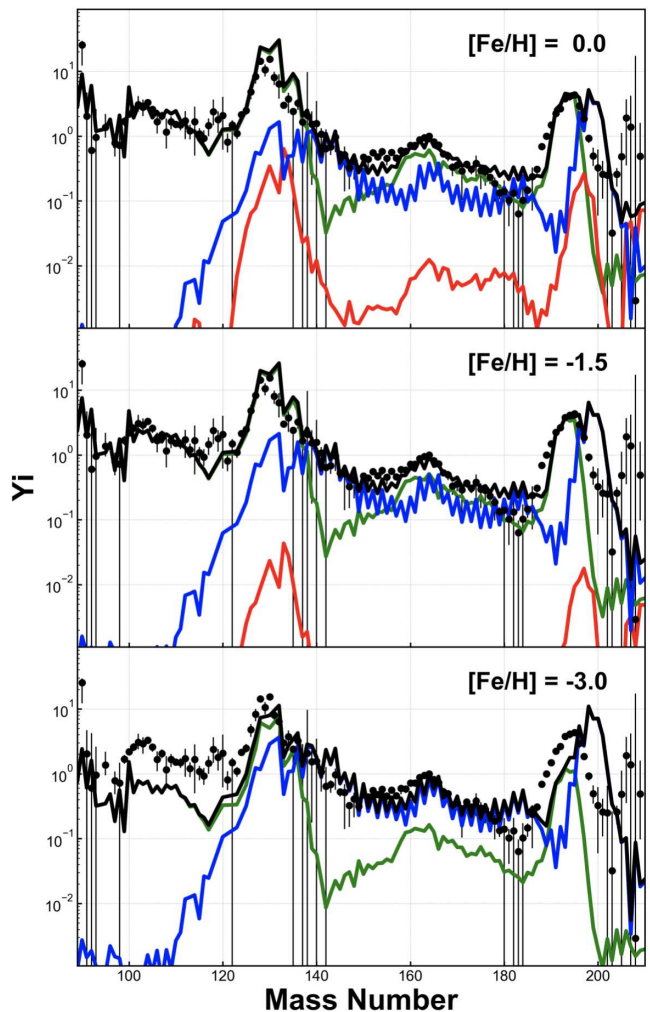}
\caption{
The time-varying isotopic abundance patterns at [Fe/H]=0.0, -1.5, -3.0 from top to bottom. The black lines show the total calculated r-process abundances.  The red and green lines represent the contributions from NSMs and CCSNe (including the NDW and MHDJ), respectively. The blue lines show the contribution from collapsars.}
\label{abund-Fe}
\end{figure}

Although model asym+ has the lowest reduced chi square, we show the fitted abundance pattern of model sym+ in the top panel of Fig.~\ref{abund-Fe} because sym+ also reproduces the metallicity evolution of europium much better than asym+.
This, however, does not necessarily mean that symmetric fission is more likely in nature. 

The abundance distribution from each of these four different r-process sites has unique characteristics particularly for nuclides heavier than the first r-process peak at A=80.
The NDW produces only light r-process elements in the mass range $A<130$. On the other hand, the MHDJ site produces r-process abundance peaks particularly around the second and third peaks at $A\approx$130 and 195. 
The CCSN sum of NDW and MHDJ is shown as  a green line on Fig.~\ref{abund-Fe}.
\begingroup
\renewcommand{\arraystretch}{1.3}
\begin{table}
\caption{\label{tab:chisq}.}
\begin{ruledtabular}
\begin{tabular}{c|rrrrrrrrrrrr}
  \ \ Model\ \ \ &\multicolumn{3}{c}{sym$-$\ \ }&\multicolumn{3}{c}{asym$-$\ \ }&\multicolumn{3}{c}{sym$+$\ \ }&\multicolumn{3}{c}{asym$+$\ \ }\\ \hline
  $\tilde{\chi}^2$&\multicolumn{3}{c}{22.3\ \ \ }&\multicolumn{3}{c}{14.5\ \ \ }&\multicolumn{3}{c}{14.1\ \ \ }&\multicolumn{3}{c}{10.8\ \ \ }
\end{tabular}
\end{ruledtabular}
\end{table}
\endgroup

Although NSMs produce r-process contributions from not only the tidal ejecta, but also from accretion disk outflows as neutrino-driven winds \cite{kajino2019}, we focus here on the dynamical tidal ejecta  as in \cite{suzuki2018} with   30 outflow trajectories based upon  the hydrodynamic simulations of~\cite{korobkin2012, piran2013, rosswog2013}. These trajectories  are based on SPH simulations in Newtonian gravity, where the neutrino transport is taken into account in a neutrino leakage scheme~\cite{rosswog2003}. 

This ejecta has a very low electron  fraction ($Y_e\sim0.1$) which causes the r-process path to run along extremely neutron-rich isotopes. The reaction flow quickly reaches heavy fissile nuclei in the region $A=250-290$. Therefore, the FFD affects strongly the final abundance distribution in the mass range of $A=100-180$ and even  heavier isotopes \cite{shibagaki2016}. When a symmetric FFD~\cite{suzuki2018} is used in the NSM r-process, the second peak around $A\sim130$ is reproduced reasonably well as shown in Fig.~\ref{abund-Fe}, while the abundance distribution is smoothed out in the model with an asymmetric FFD~\cite{shibagaki2016}. Nevertheless, there are several common features such as only intermediate-to-heavy mass isotopes having $A>100$ are produced,  and  the third peak is slightly shifted towards the heavier mass region since the neutron-number density remains high after freezeout in tidal ejecta from NSM.

The abundance pattern in the collapsar r-process (blue line in Fig.~\ref{abund-Fe}) has three unique features due to the very rapid-neutron captures caused by the relatively high values of the neutron density at freezeout~\cite{famiano2020}. First, the r-process peaks are shifted systematically towards the heavier mass region. This is caused by residual neutron captures after freezeout of the r-process.   This causes a discrepancy away from the observed solar r-process abundance pattern around the third peak $A\sim195$ (Fig.~\ref{abund-Fe}).
Secondly, an odd-even pattern manifests in the lanthanide abundance hill.  This is a typical feature resulting from very rapid-neutron captures and the sudden freezeout. We note, however, that the  calculation of Ref.~\cite{siegel2019} adopts  a different ejecta model and shows a smoother pattern.
Thirdly, the second peak in the collapsar model is broadened to higher mass region A$=140-150$ as mentioned previously.

The properties of neutron-rich unstable nuclei and the physical conditions of the trajectory dynamics are thus intricately connected  in any nucleosynthesis simulations. However, if some characteristic features, such as those  discussed here, were measured in the isotopic abundances of r-process enhanced metal-deficient stars, this could confirm that the heavy nuclei in those stars originate from a single or a few very similar r-process events.

Our simple GCE model indicates a time-metallicity relation given by t/10$^{10}$y $\approx$ $10^{[{\rm Fe}/{\rm H}]}$.  However,  this relation is broken in the extremely metal-deficient region [Fe/H]$<$-2.  This is  because of the inhomogeneous nature of the  stochastic Galactic star formation~\cite{hirai2015} at low metallicity.  Nevertheless, metallicity remains  a reasonable measure of the time evolution of the Galaxy down to  -2$<$[Fe/H]. 

Figure~\ref{abund-Fe} shows the abundances of r-process nuclei from each contribution from NDW, MHDJ, NSMs and collapsars as snapshots in metallicity or time. Among these possible astrophysical sites, CCSNe (i.e. NDW and MHDJ) and collapsars make the predominant contribution in the early Galaxy. The NSM contribution grows gradually with increasing metallicity and eventually reaches 1 \% of the total abundance of solar r-process elements at [Fe/H] = -1.5, -1.3, and -0.7 for models with a minimum coalescence times of $\tau_g$= 1, 10, and 100 My, respectively.
We thus conclude that the NSM contribution was negligibly small in the early Galaxy and has arrived later in Galactic evolution due to the long coalescence time delay $\tau_g$. This conclusion is nearly independent of the model selection of astrophysical sites and input nuclear physics. 

However, the detailed time variation of the abundance pattern does depend on the  input nuclear physics or models of the ejecta used in r-process simulations. 
A typical example of the dependence on input nuclear physics is the FFD. As discussed previously, 
symmetric and asymmetric FFDs can lead to very different abundance patterns over the  entire mass range. Although the MHDJ r-process (green line) explains well the abundance peaks around $A\sim$ 130 and 195 and the hill of lanthanides around $A\sim165$, a deficiency of isotopes in $A=140\sim150$  remains in either FFD model. Indeed, most r-process nucleosynthesis calculations underproduce the heavier isotopes just above the second peak \cite{kajino2019}. 

In Model sym+, NSMs change the total abundance pattern only slightly when the Galaxy has evolved to near solar metallicity [Fe/H]=0. This is because the NSM contribution fraction is only about 1\%. In the metal-deficient region [Fe/H]$<-1.5$, the NSM contribution does not change the total abundance pattern because the r-process nuclei are dominated by CCSNe or collapsars.
On the other hand, the abundance pattern changes drastically as a function of metallicity at  [Fe/H] $<-1.5$.  Figure~\ref{abund-Fe} exhibits a very busy abundance pattern because of the odd-even structure in the collapsar r-process yields. There are  significant shifts of the second and third peaks towards heavier mass numbers as shown by the black lines.

\begin{figure}
\includegraphics
[width=1.0\linewidth]
{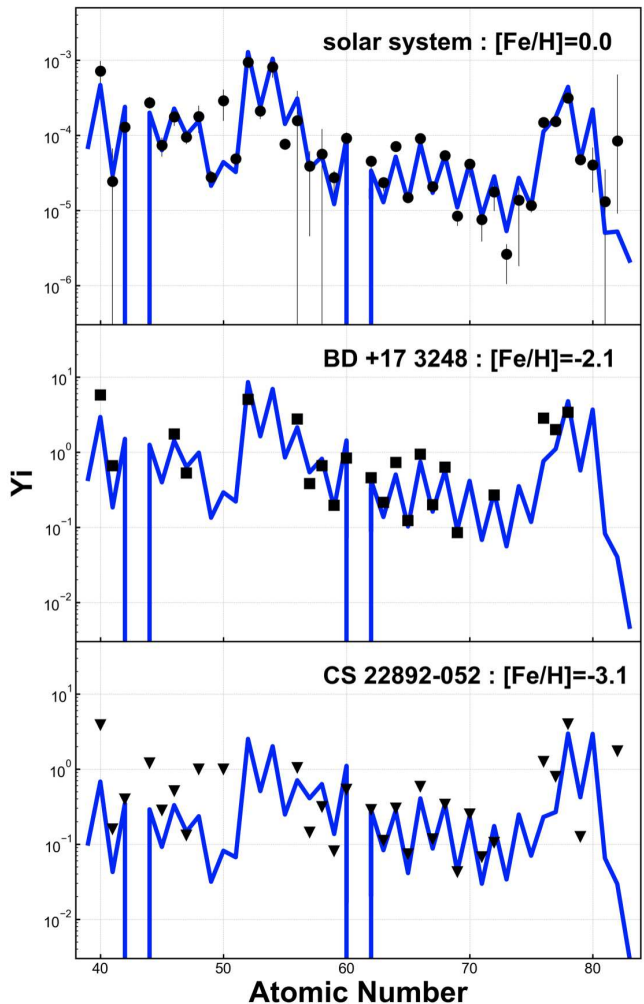}
\caption{\label{fig:elemental} Comparison of  calculated r-process elemental abundance patterns (blue lines) with the observed abundances (points)  of the solar system (top) and metal poor stars, BD+173248 (middle) and CS 22892-052(bottom).}
\end{figure}


Figure~\ref{fig:elemental} displays the calculated elemental abundance patterns as a function of atomic number $Z$ in Model sym+, compared with observational data in the r-process enhanced metal-poor halo stars, BD+173248 ([Fe/H] = -2.1)~\cite{roederer2012} and CS22892-052 ([Fe/H] = -3.1)~\cite{sneden2003}.  

They exhibit a more or less similar elemental abundance pattern for any metallicity, except for several atomic numbers to be discussed below. In particular, around the lanthanide hill near Dy ($Z=66$), they agree with each other independently of any models.  This feature is known as the universality of the r-process elemental abundance pattern~\cite{sneden2008}. Such similarity is due to the fact that there are many isotopes contributing to the same atomic number $Z$ with different mass numbers $A$~\cite{shibagaki2016}. The most abundant isotopes smooth the detailed structure apparent in the mass distributions of Fig.~\ref{abund-Fe}.

The peak height around Te ($Z=52$) and Os ($Z=78$) depends on the models. This model dependence arises from the different contribution fractions from the four astrophysical sites.
Unfortunately, the second peak elements except for Te~\cite{roederer2012} have not been observed in metal-deficient halo stars.


 The elemental abundances are in reasonable agreement with observational data.
Interestingly, the actinides also are remarkably enhanced in the collapsar r-process. This is because the extremely high neutron number density in collapsars causes the  r-process path to proceed along very neutron-rich nuclei and produce  neutron-rich isotopes beyond the third peak. Such a remarkable enhancement of actinides is indeed observed in actinide-boost stars~\cite{Mashonkina2014}.

One finds a discrepancy for lighter elements Zr-Sn ($Z=38-48$) in CS 22892-052. A remarkable enhancement of these elements has been reported in so-called Honda stars~\cite{honda2006}. Our present theoretical interpretation is that the universality between the second and third peaks including the lanthanide hill around $Z\approx66$ and beyond is satisfied in any cases, although the variation in a wider mass range can be reasonably explained by inhomogenity in the early galaxy. 


To summarize, we have studied the cosmic evolution of the r-process abundance pattern in the context of  GCE models that take into  account multiple astrophysical sites simultaneously  (i.e. NDW and MHDJ CCSNe, NSMs and collapsars). The NSM r-process calculations were carried out with different input nuclear physics including  symmetric and asymmetric FFDs. We then find that the r-process elements in the early Galaxy are dominated by the yields from CCSNe and collapsars, while the NSM contribution is inevitably delayed due to the cosmologically long coalescence timescale for very slow GW radiation. The relative NSM contribution rapidly increases with cosmic time. However, it does not reach even 1\% of the total solar r-process composition until the metallicity is enriched to [Fe/H]$\ge$ -1.5.
This conclusion does not change for a wide range of minimum coalescence times $\tau_g$ = 1 - 100 My in any GCE models including multiple sites and different input nuclear physics.

We also find that significant differences among 
our multiple-site GCE model calculations arise in the isotopic abundance pattern as a function of  mass number $A$, while still satisfying the universality of elemental abundances for metal-poor halo stars.
This is in contrast to previous studies that focused on only a single r-process site or a combination of at most two astrophysical sites in order to explain the universality of the elemental abundance pattern. 

Several unique features of each astrophysical site are still expected in the GCE of the isotopic mass $A$-abundance pattern. In particular, the collapsar  contribution dominates from the very beginning of the early Galaxy since its progenitor is a very massive star. The collapsar r-process shows an odd-even pattern over the entire mass range and also both the collapsar and NSM the abundance peaks shift towards the heavier mass region  due to a high residual neutron-flux during the freezeout of the r-process.

Although the elemental $Z$-abundance patterns are more or less similar to one another among the models, one can find exceptional differences in the actinides or light r-process elements Z $<$ 42. Therefore, these are the important indicators of the  dominant r-process site. Also, the peak structure is model dependent  due to the different contribution fractions from the four astrophysical sites considered here.  

It is therefore highly desirable to carry on spectroscopic observations with  next generation telescopes such as the Thirty Meter Telescope~\cite{tmt}.  These could provide the metallicity dependence of the abundance ratios of actinides, lanthanides and lighter elements as well as the abundance peaks simultaneously. A separation of each r-process element into isotopes should provide constraints on the  evolution of the NSM, MHDJ and collapsar  model construction beyond that of a single r-process site as discussed in this article. 

\begin{acknowledgments}
This work is supported in part by Grants-in-Aid for Scientific Research of JSPS (20K03958, 17K05459).
Work at the University of Notre Dame supported by DOE nuclear theory grant DE-FG02-95-ER40934. MAF is supported by NASA grant 80NSSC20K0498.
\end{acknowledgments}

\nocite{*}

\bibliographystyle{apsrev4-2}
\bibliography{aapmsamp}

\end{document}